\documentclass[useAMS,usenatbib]{mn2e}
\usepackage{amssymb,amsmath,epsfig,times, natbib}
\title[The ICM around H1821+643]{The effect of the quasar H1821+643 on the surrounding intracluster medium: revealing the underlying cooling flow }
\author[S. A. Walker et al.]{S. A. Walker,$^1$\thanks{Email: 
    swalker@ast.cam.ac.uk} A. C. Fabian$^1$, H. R. Russell$^1$ and J. S. Sanders$^2$\\
  $^1$Institute of Astronomy, Madingley Road, Cambridge CB3 0HA \\
    $^2$Max-Planck-Institute fur extraterrestrische Physik, 85748 Garching, Germany \\
  \\
    \\
   \\
   \\
}
\date{}

\begin{document}

\maketitle

\begin{abstract}
We present a detailed study of the thermodynamic properties of the intracluster medium of the only low redshift galaxy cluster to contain a highly luminous quasar, H1821+643. The cluster is a highly massive, strong cool core cluster. We find that the ICM entropy around the quasar is significantly lower than that of other similarly massive strong cool core clusters within the central 80 kpc, and that the entropy lies significantly below the extrapolated baseline entropy profile from hierarchical structure formation. By comparing the scaled temperature profile with those of other strong cool core clusters of similar total mass, we see that the entropy deficiency is due to the central temperature being significantly lower. This suggests that the presence of the quasar in the core of H1821+643 has had a dramatic cooling effect on the intracluster medium around it. We find that, if the quasar was brighter in the past, Compton cooling by radiation from the quasar may have caused the low entropy and temperature levels in the ICM around the quasar. Curiously, the gradients of the steep central temperature and entropy decline are in reasonable agreement with the profiles expected for a constant pressure cooling flow. It is possible that the system has been locked into a Compton cooled feedback cycle which prevents energy release from the black hole heating the gas sufficiently to switch it off, leading to the formation of a huge ($\sim$3$\times$10$^{10}$ M$_{\odot}$) supermassive black hole.  
 
\end{abstract}


\section{Introduction}

H1821+643 is the only low redshift ($z=0.299$), highly luminous quasar to reside in the centre of a galaxy cluster. The reason why this galaxy cluster harbours a quasar remains unknown. X-ray observations of the intracluster medium (ICM) around this quasar provide the best means of understanding the interaction between the quasar and the surrounding galaxy cluster. It is the only galaxy cluster for which a detailed spatially resolved analysis of the thermodynamic properties of the ICM (temperature and entropy) surrounding a highly luminous quasar can be made.  

A detailed study of the Chandra observations of H1821+643 was conducted in \citet{Russell2010}, which details all of the data reduction and spectral analysis. Here we extend the study of the intracluster medium around H1821+643 by exploring the entropy distribution and comparing it with simulations and other massive cool core clusters which are similar to H1821+643. Studying the entropy profile is a powerful tool for understanding the intracluster medium as it contains a record of the past cooling and heating. Specific predictions have been made for the entropy level expected from hierarchical structure formation assuming only gravitational processes (\citealt{Voit2005}), allowing us to gain an insight into the non-gravitational heating and cooling which the ICM has experienced. However as heating and cooling raise and lower the entropy repectively, cancellation between these two may occur.  

Any deviation from this baseline entropy profile must be due to non-gravitational processes. Typically an entropy excess over the baseline profile is observed, which is greater for less massive clusters and groups (\citealt{Sun2009}, \citealt{Pratt2010}, \citealt{Walker2013_Centaurus}, \citealt{McDonald2014}), due to the non-gravitational energy input into the ICM (through AGN-feedback and supernovae for instance) having a greater impact on gas in the shallower gravitational potential wells of less massive clusters. The cluster containing the quasar H1821+643 is a highly massive, relaxed cool core system, ($M_{500}=9 \times 10^{14}$ M$_{\odot}$)\footnote{$M_{500}$ is the mass within a radius $r_{500}$ within which the mean density is 500 times the critical density of the universe at redshift of the cluster.} , and would be expected to have only a very small entropy excess above the baseline profile. For example, \citet{Morandi2007} used Chandra to study the entropy profiles of very massive, strong cool core clusters (A2204, A2390, A1835, Zw3146, MS1358.4+6245 and RXJ1347.5−1145), similar to H1821+643, and found that they are in good agreement with the Voit baseline profile down to 0.03$r_{2500} \simeq 0.01r_{500}$ with a very low level of scatter. This indicates that in these strong cool core clusters the effects of heating and cooling on the entropy profile have cancelled out.

\begin{figure}
  \begin{center}
    \leavevmode
    \hbox{
      \epsfig{figure=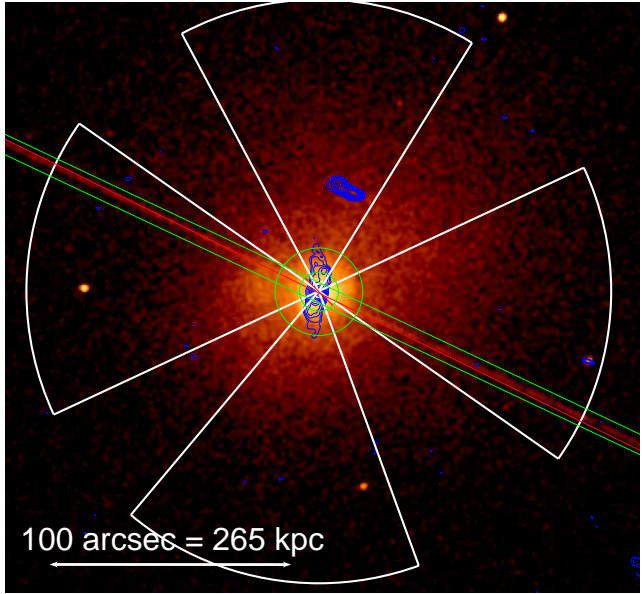,
        width=\linewidth}
            }
   
      \caption{Chandra image of the quasar and the surrounding intracluster medium in the 0.7-7.0 keV band. Overplotted in blue are the 1.4 GHz VLA radio contours from \citet{Blundell2001}, showing the jets to the north and south. The central 8 arcsecs around the quasar and the readout streak are masked out to ensure we only study the intracluster medium of the surrounding cluster. The region between the two green circles is the 30-80 kpc in which the temperature and entropy profile steepen dramatically, as shown in Fig. \ref{T_deproj_compare}. The white sectors show the regions along the jet directions to the north and south, and away from the jet direction to the east and west, which we compare in section \ref{cooling_flow_comparison}.  }
      \label{image}
  \end{center}
\end{figure}

\citet{Russell2010} however found that the entropy (and temperature) profile shape changed significantly around 80 kpc ($0.08r_{500}$) from the core of the cluster, becoming far steeper than the baseline entropy profile between 30-80 kpc (8-20 arcseconds) from the core. Here we perform a detailed study to understand this entropy behaviour, and the impact the central quasar may have had on its surrounding ICM. The Chandra image of the quasar and the surrounding intracluster medium in the 0.7-7.0 keV band is shown in Fig. \ref{image}. In all of the analysis the central 8 arcseconds (30kpc) around the quasar are neglected, shown by the inner circle in Fig. \ref{image}, as well as the readout streak. Simulations of the quasar surface brightness profile conducted in \citet{Russell2010} show that outside 8 arcseconds the cluster emission is dominant, and the emission from the quasar is negligible. The 30-80 kpc region where the entropy profile and temperature profiles steepen is shown by the two circles in Fig. \ref{image}.

 We use a standard $\Lambda$CDM cosmology with $H_{0}=70$  km s$^{-1}$
Mpc$^{-1}$, $\Omega_{M}=0.3$, $\Omega_{\Lambda}$=0.7. All errors unless
otherwise stated are at the 1 $\sigma$ level. 

\section{Comparing entropy and temperature profiles}

We use the deprojected temperature, density and entropy profiles obtained in \citet{Russell2010} using an X-ray surface brightness deprojection code (which includes additional assumptions of hydrostatic equilibrium and an NFW total mass model). This allows finer spatial binning than a direct spectral deprojection method (this method was first described in \citealt{Fabian1981} and also used in, for example, \citealt{White1997} and \citealt{Allen2001}). This is discussed in detail in section 4.3 of \citet{Russell2010}. 

\begin{figure}
  \begin{center}
    \leavevmode
    \hbox{
      \epsfig{figure=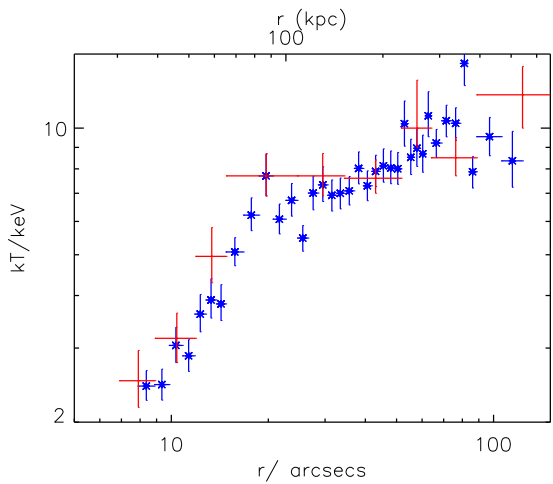,
        width=\linewidth}
            }
   
      \caption{Comparing the temperature profile obtained using the surface brightness deprojection method from \citet{Russell2010} (blue points) with direct spectral deprojection using \textsc{dsdeproj} with finer binsize (red points), showing the two profiles to agree well with one another.     }
      \label{T_deproj_compare}
  \end{center}
\end{figure}

To further check that the temperature and entropy profiles obtained using the surface brightness deprojection method are not affected by the need to assume an NFW profile for the total mass profile, we also found the deprojected temperature profile using the \textsc{dsdeproj} direct spectral deprojection method (\citealt{Sanders2007}, \citealt{Russell2008}). As this requires spectral fitting, the annuli widths are necessarily larger than for the surface brightness deprojection method. An original comparison was made in \citet{Russell2010}, showing the two profiles to be in good agreement, but here we divide the central 2 bins originally used in \citet{Russell2010} into four bins to improve the spatial resolution and test the sharp temperature drop between 8-20 arcsecs. The comparison is shown in Fig. \ref{T_deproj_compare}, where we see that the two profiles agree well, indicating the the sharp change in temperature profile gradient is robust, and not a consequence of assuming a mass model. In all regions, the deprojected spectra are well fit by a single absorbed apec component, and the fit is not improved by adding second temperature component or a powerlaw component. 

\begin{table}
  \begin{center}
  \caption{Sample of galaxy clusters used for comparisons listed in mass order. All of these are cool core clusters except for A1650, A2142 and A2244. }
  \label{Cluster_sample}
  
    \leavevmode
    \begin{tabular}{lllllll} \hline \hline
    Cluster &z&  $M_{500}$/10$^{14}$ $M_{\odot}$\\ \hline
A262 &     0.016 &       1.4 \\
AWM7 &     0.017 &       2.6 \\
A2597 &     0.085 &       2.7 \\
A1068 &      0.138 &       2.9 \\
RXCJ1542.2-3154 &      0.103 &       3.6 \\
RXCJ1558.3-1410 &     0.097 &       3.8 \\
A1664 &      0.128 &       3.8 \\
A1204 &      0.171 &       3.9 \\
A2244 &     0.097 &       4.9 \\
A1795 &     0.062 &       4.9 \\
A1650 &     0.084 &       5.2 \\
RXCJ1459.4-181 &      0.24 &       6.0 \\
RXCJ1023.8-2715 &      0.253 &       6.5 \\
RXCJ1504.1-0248 &      0.215 &       6.8 \\
A2029 &     0.077 &       7.1 \\
A478 &     0.088 &       7.1 \\
Zw3146 & 0.291 &       7.1 \\
A2142 &     0.091 &       8.3 \\
A1835 &      0.253 &       8.8 \\
A2390 & 0.232 &       10.3 \\
A2204 & 0.152 &       11.5 \\
RXJ1347-1145 & 0.451 &       14.9 \\ \hline
H1821+643 & 0.299 & 9.0 \\ 
      \hline

    \end{tabular}
  \end{center}
\end{table}

To further ensure the robustness of our study of the ICM, we very conservatively neglect the central 8 arcseconds (the inner 30kpc) around the central quasar. Simulations of the quasar surface brightness profile performed in \citet{Russell2010} show that the cluster emission outside 8 arcseconds from the quasar is completely dominant, and the quasar contribution is negligible (see figure 7 of \citealt{Russell2010}). 

To compare the cluster with other clusters, we use the cluster sample compiled in \citet{Russell2010}, and which is tabulated in Table \ref{Cluster_sample}. These clusters were selected to provide a large mass range of cool core clusters with similar spatial extent on the ACIS chip to H1821+643, and to ensure that at least 8 deprojected radial bins could be made so that the spatial resolution was roughly the same between the clusters. Three non-cool core clusters are included to provide a contrast for the entropy and temperature profiles (these are A1650, A2142 and A2244). We also reanalyse the strong cool core clusters studied in \citet{Morandi2007} and add these to the comparison sample, (these are A1835, A2204, Zw3146, A2390 and 
RXJ1347-1145), which are in the same mass range as the cluster underlying H1821+643 (we omit MS1358.4+6245 as the data quality for this cluster is significantly lower than that of the other clusters). All of the deprojected temperature, density and entropy profiles for these clusters were obtained using \textsc{dsdeproj} following the methods described in \citet{Russell2010}.

First, we compare the entropy ($K=kT/n_{e}^{2/3}$) profiles with the baseline profile of \citet{Voit2005}
\begin{equation}
K(R)/K_{500} = 1.47(r/r_{500})^{1.1}.
\label{eqn:KR}
\end{equation}
which is the entropy profile obtained from numerical simulations of hierarchical structure formation in which only gravitational processes are considered. The entropy profiles for each cluster are scaled appropriately by the expected self similar entropy at $r_{500}$ 
\begin{eqnarray}
\label{eqn:K500}
K_{500} &=& 106 \ {\rm keV\ cm}^{-2} \left( \frac{M_{500}}{10^{14}\,h_{70}^{-1}\,M_\odot} \right)^{2/3}\, \left(\frac{1}{f_b}\right)^{2/3}\, \nonumber \\ 
&& \times E(z)^{-2/3}\, h_{70}^{-4/3}
\end{eqnarray}
using $f_{b}$=0.15 as in \citet{Pratt2010}. To obtain the cluster masses, we use a method similar to that used in \citet{Schmidt2007}. This involves using the deprojected density profiles to predict the temperature profile assuming hydrostatic equilibrium and the NFW profile for the total mass, moving inwards from the outermost annulus (the results are unchanged if the method starts from the innermost annulus and moves outwards). The total mass profile and the gas mass fraction profile for the cluster surrounding H1821+643 are shown in Fig. \ref{massprofile} and Fig. \ref{fgasprofile} respectively in Appendix \ref{sec:appendix}.

To check the mass determination for the cluster surrounding H1821+643, we compare with the mass expected using the $M_{500}-Y_{X}$ relation of \citet{Arnaud2007}. We can trace the surface brightness profile out to $r_{500}$, and determine that the total gas mass mass within $r_{500}$ is $M_{g,500}=1.3\times10^{14}$ M$_{\odot}$. The core excluded spectroscopic temperature (within 0.15-0.75$r_{500}$) is 8.9 keV. The $M_{500}-Y_{X}$ relation of \citet{Arnaud2007} yields $M_{500}=8.9 \times 10^{14}$ M$_{\odot}$, in good agreement with the measurement of $9.0 \times 10^{14}$ M$_{\odot}$ using the method of \citet{Schmidt2007}.

\begin{figure*}
  \begin{center}
    \leavevmode
    \hbox{
      \epsfig{figure=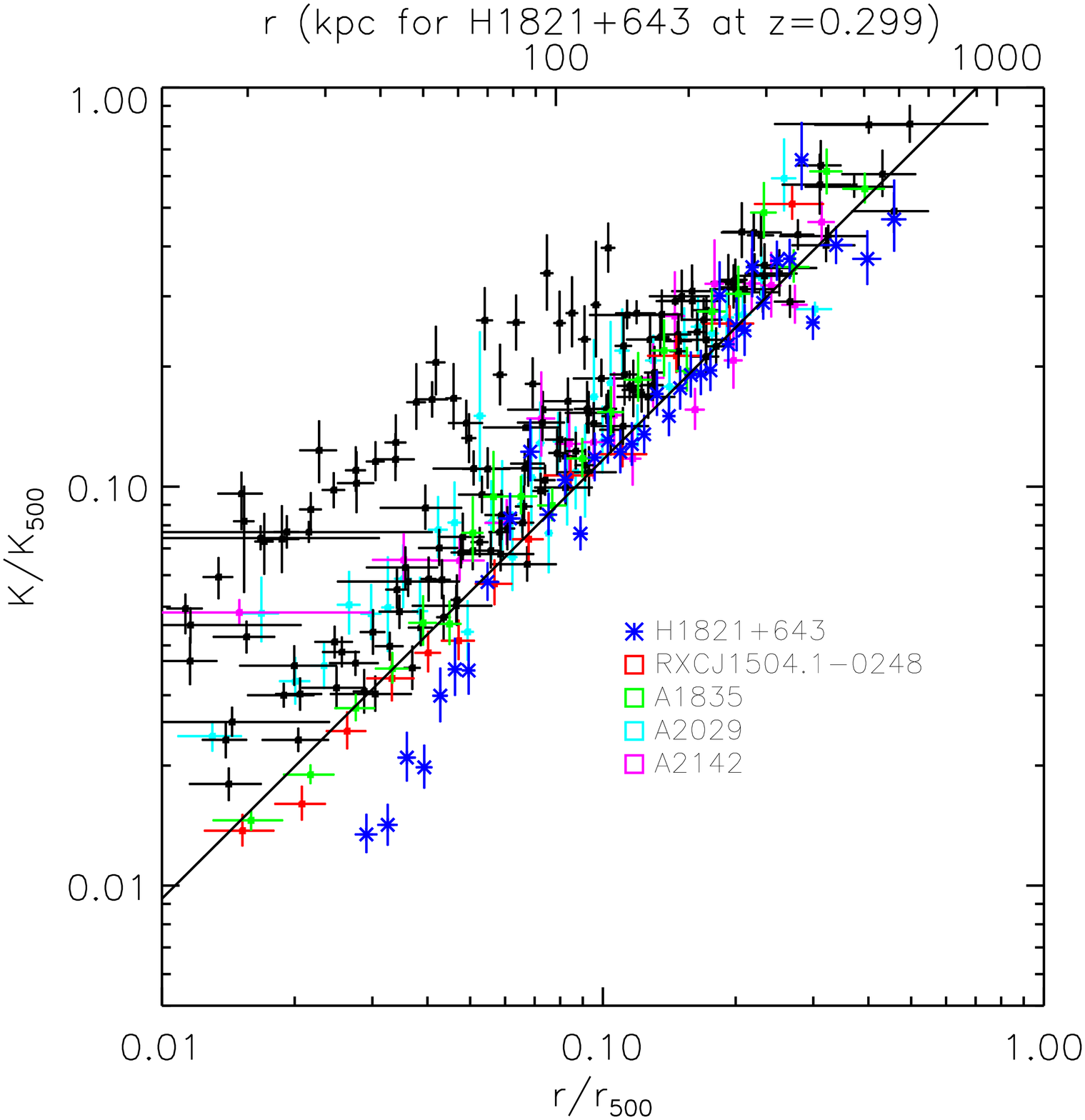,
        width=0.45\linewidth}
         \epsfig{figure=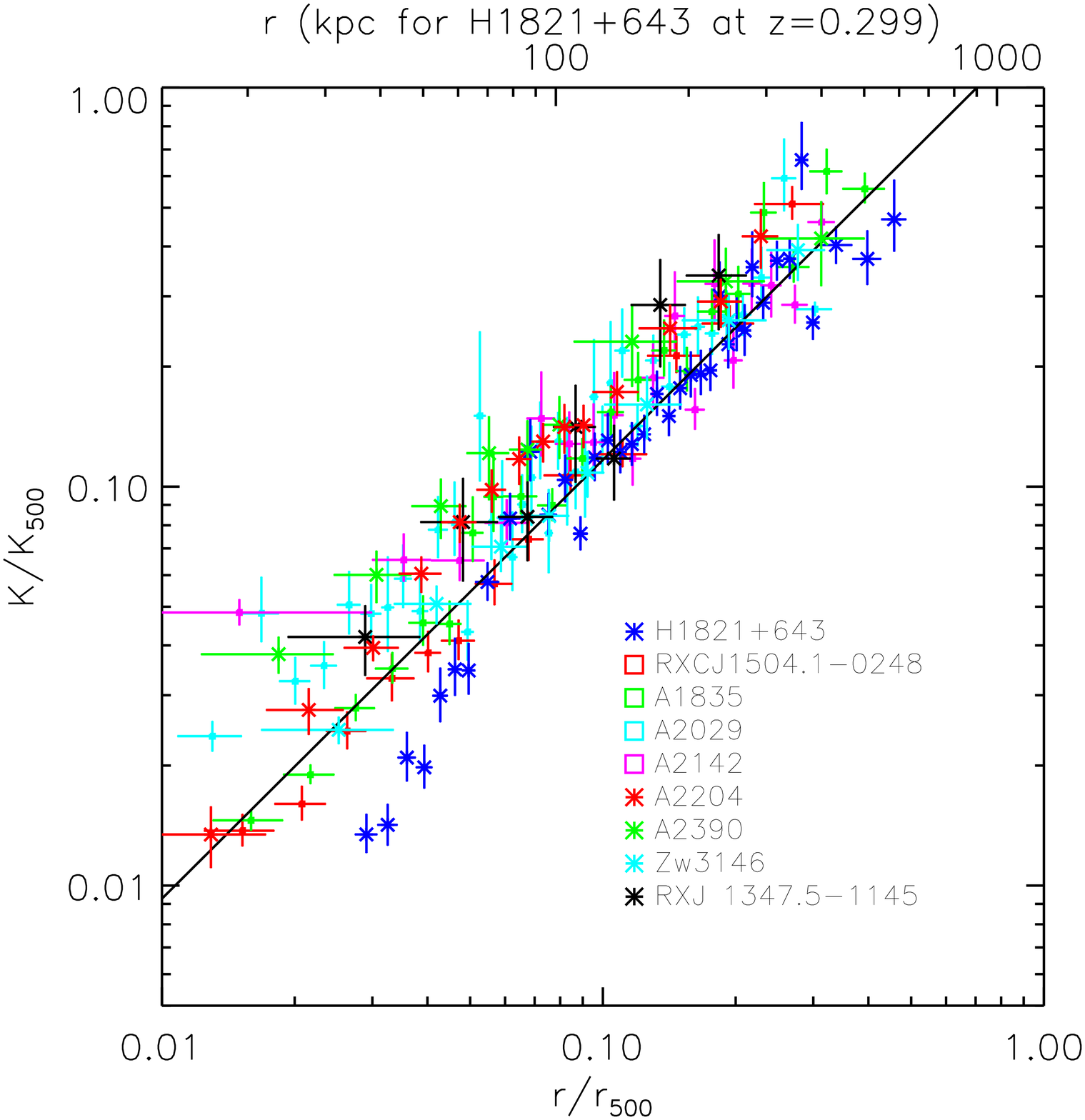,
        width=0.45\linewidth}     
        }
   
      \caption{\emph{Left}: Entropy profiles for the sample of clusters studied in \citet{Russell2010} with the expected self similar evolution scaled out, and the most massive clusters shown in colour and labelled. All of the other, lower mass, clusters in the \citet{Russell2010} sample are plotted in black and demonstrate an entropy excess over the baseline profile as expected. \emph{Right}: The scaled entropy profiles for just the most massive cool core clusters in the sample. The solid black line shows the powerlaw baseline entropy profile from \citet{Voit2005}.    }
      \label{K_profiles}
  \end{center}
\end{figure*}

Fig. \ref{K_profiles} shows the scaled entropy profiles for the sample of clusters studied in \citet{Russell2010} compared to the Voit baseline profile, with the left hand panel showing all of the clusters, and the right hand panel showing only the most massive clusters which are would be expected to be similar to H1821+643. We see that the entropy profile for H1821+643 has a remarkable drop compared to all of the other clusters within 0.07$r_{500}$ (80kpc for H1821+643), with its entropy lying 
significantly below the baseline level, and having a far steeper gradient than the $r^{1.1}$ baseline powerlaw. This indicates that the ICM in the central 80 kpc of H1821+643 has undergone significant cooling. 

\begin{figure*}
  \begin{center}
    \leavevmode
    \hbox{
      \epsfig{figure=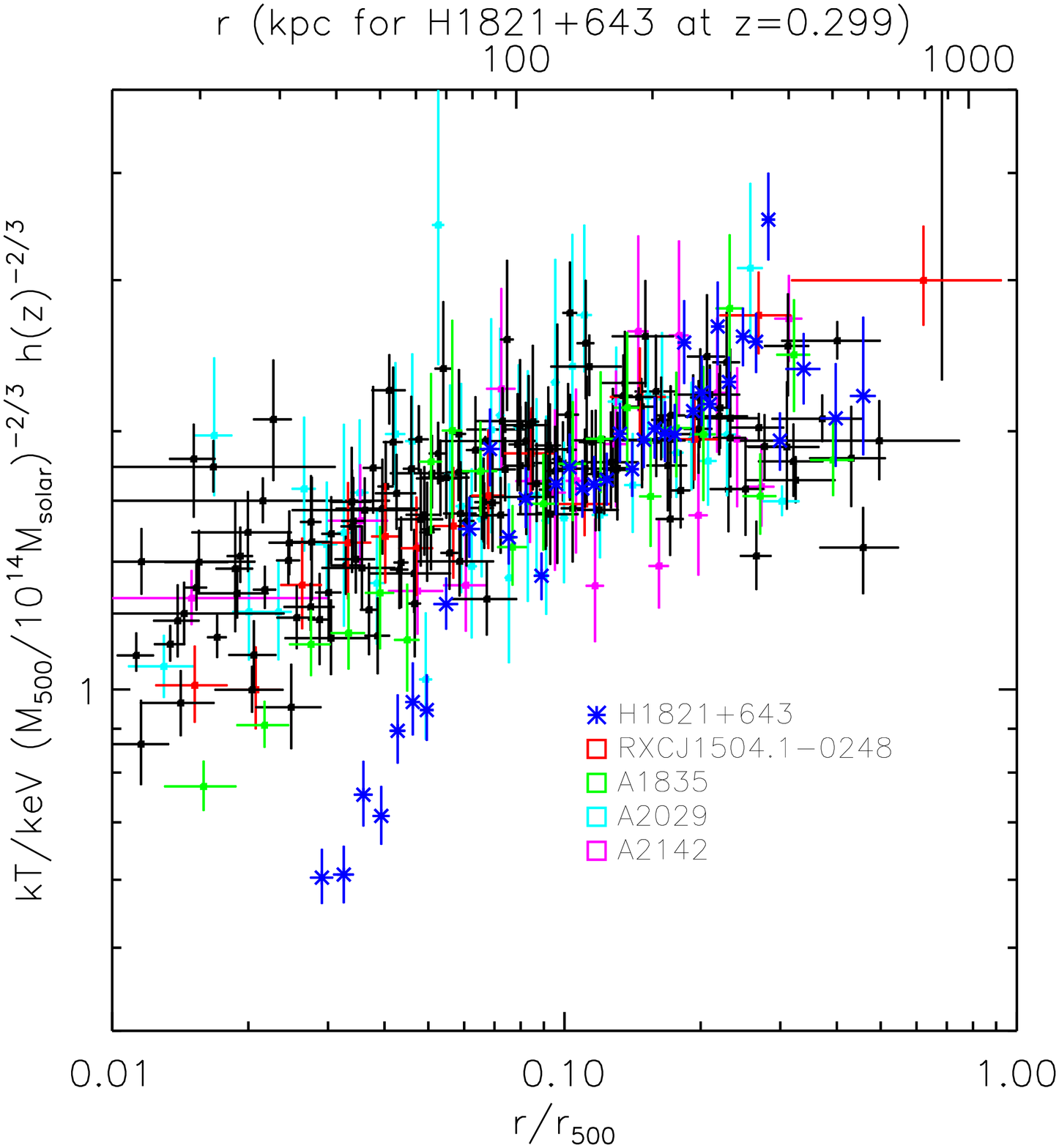,
        width=0.45\linewidth}
         \epsfig{figure=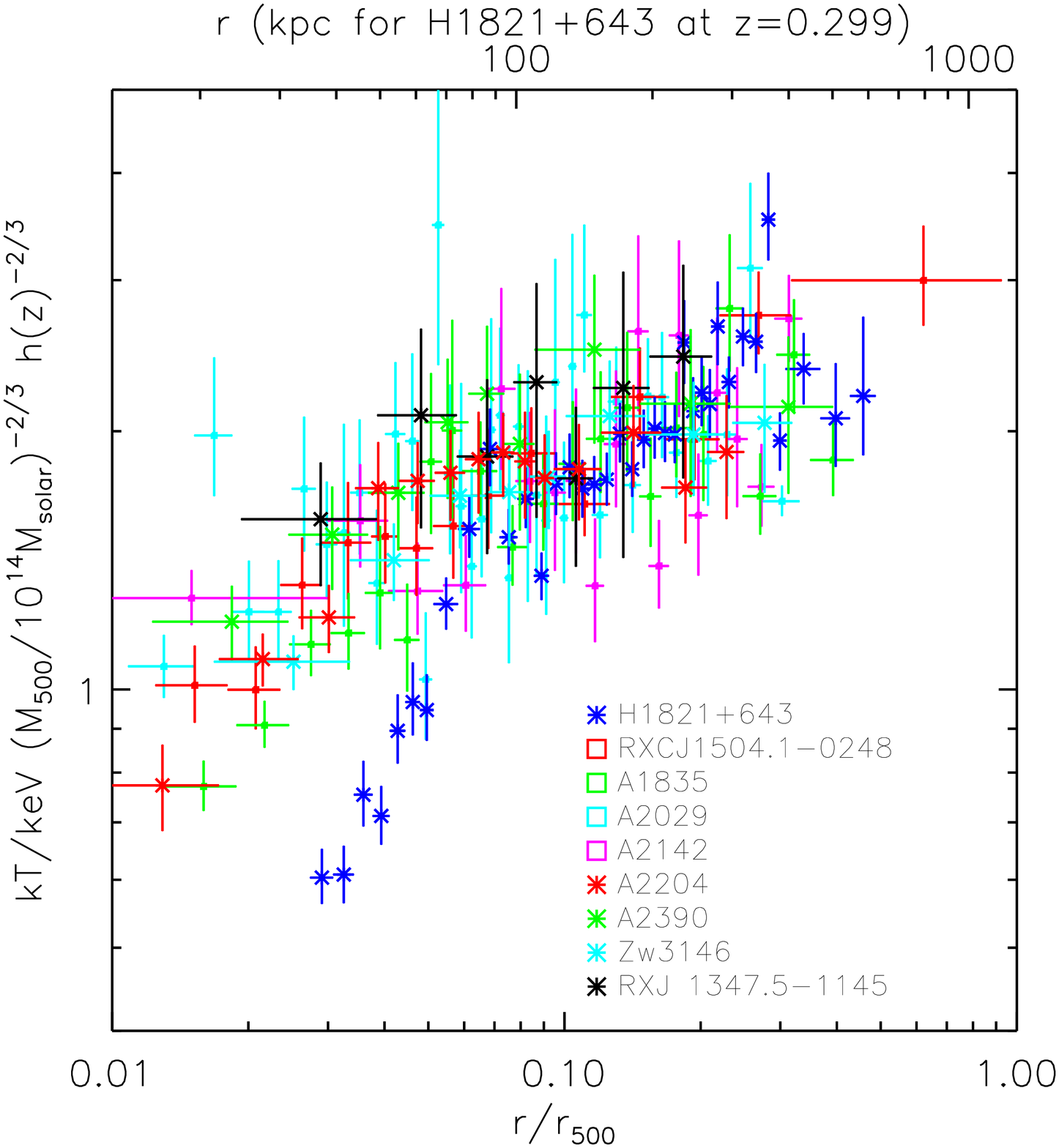,
        width=0.45\linewidth}     
        }
   
      \caption{\emph{Left}:Temperature profiles for the sample of clusters studied in \citet{Russell2010} with the expected self similar evolution scaled out, and with the most massive clusters shown in colour and labelled. All of the lower mass clusters from the \citet{Russell2010} sample are plotted in black. \emph{Right}: The scaled temperature profiles for the most massive cool core clusters in the sample.     }
      \label{T_profiles}
  \end{center}
\end{figure*}

To investigate the cause of the entropy deficit in the core of H1821+643, we next examine the scaled deprojected temperature profiles. In Fig. \ref{T_profiles} we compare the temperature profiles of the sample of clusters studied in \citet{Russell2010}, with the expected self similar evolution due to mass and redshift ($T \propto M^{2/3}h(z)^{2/3}$) scaled out. The left hand panel shows the entire sample which spans a large range of cluster mass. The right hand panel shows only the most massive cool core clusters which are most similar to the cluster containing H1821+643, and which should have similar scaled temperature and entropy profiles. We see that within 0.07$r_{500}$ the temperature profile is far steeper than all of the other clusters, and this sharp drop in temperature has caused the entropy profile to steepen and be lower than the other clusters.

\subsection{Correcting the entropy for gas depeletion}
\begin{figure}
  \begin{center}
    \leavevmode
    
      \epsfig{figure=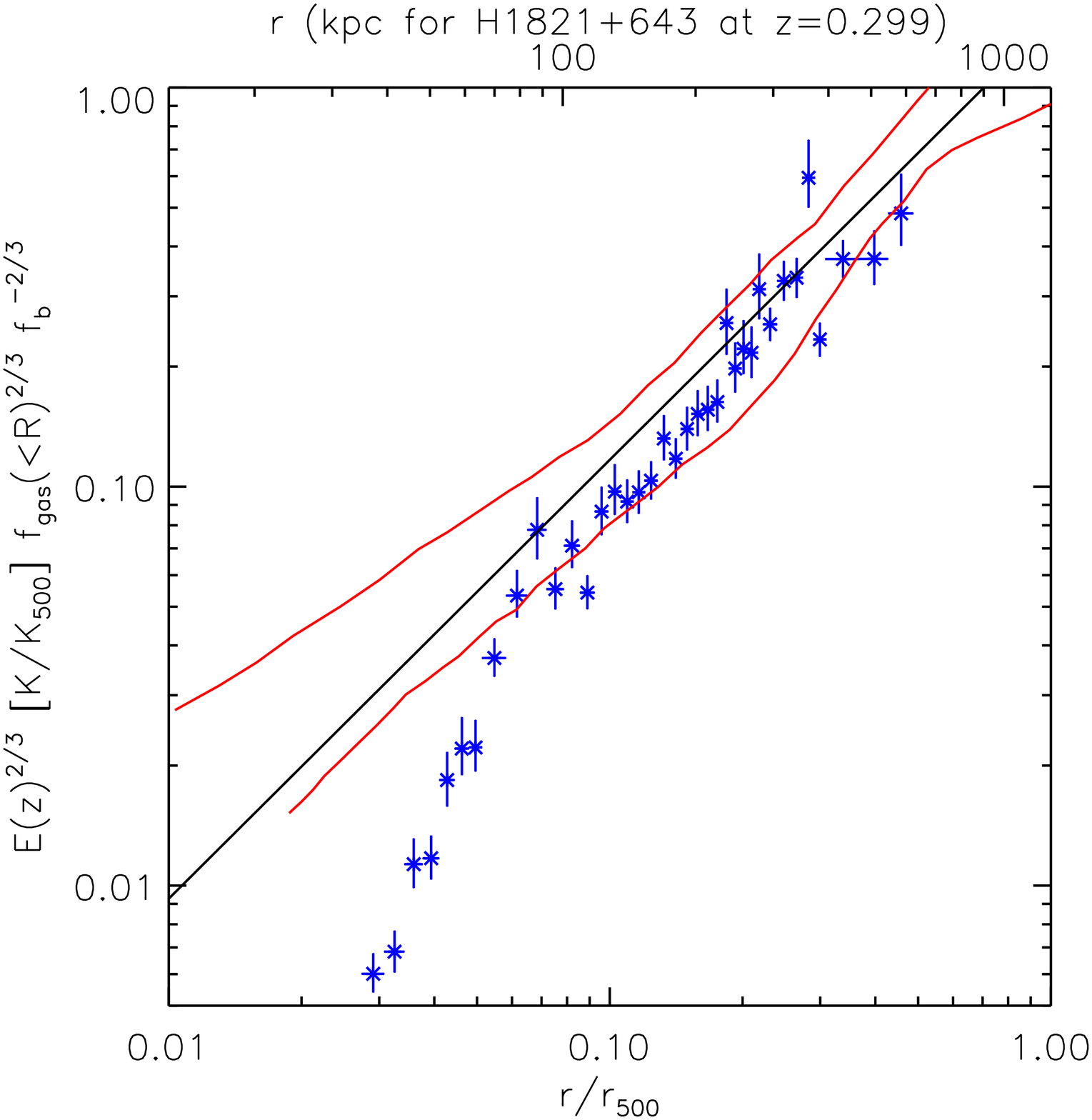,
        width=0.9\linewidth}
   
      \caption{Scaled entropy profiles mulitplied by the gas mass fraction profile as in \citet{Pratt2010}. The range of the $f_{gas}$ corrected entropy profiles from the REXCESS clusters is shown by the solid red lines, and the baseline entropy profile is the solid black line. We see that the central $f_{gas}$ corrected entropy for H1821+643 is significantly below that of the baseline entropy profile and the REXCESS clusters.    }
      \label{fgascorrectedK}
  \end{center}
\end{figure}

Entropy modification leaves both a radial and mass dependence in the gas mass fraction as gas is either removed or prevented from accreting. Multiplying by the gas mass fraction profile acts to simultaneously correct for the global and radial dependence of the gas mass fraction, and \citet{Pratt2010} found that performing this correction for the REXCESS clusters brought their scaled entropy profiles into good agreement with the baseline entropy profile, and significantly reduced the scatter.   
In this section, we follow \citet{Pratt2010} and investigate the effect of correcting the dimensionless entropy profile by multiplying by the gas mass fraction profile, which is shown in Fig. \ref{fgascorrectedK}. We see that the rescaled entropy profile remains significantly below the baseline entropy profile in the central 80 kpc. The range of $f_{gas}(<r)$ corrected entropy profiles from the REXCESS clusters is shown by the solid red lines in Fig. \ref{fgascorrectedK} (taken from figure 9 of \citealt{Pratt2010}), and we see that the $f_{gas}(r)$ corrected entropy for H1821+643 lies significantly below this range in the central 80kpc.

%\begin{table}
%  \begin{center}
%  \caption{Observational parameters of the pointings}
%  \label{obsdetails}
%  
%    \leavevmode
%    \begin{tabular}{llllll} \hline \hline
%    Obs. ID & Observatory & Total clean exposure   \\ 
%         &  & per detector (ks)  \\ \hline
%    11757 & Chandra & 20  \\ 
%      0201902201  & XMM-Newton & 21.5  \\ \hline
%
%    \end{tabular}
%  \end{center}
%\end{table}

\section{Discussion}  
\subsection{Compton cooling}

One exciting possibility is that Compton cooling from the strong UV radiation emitted by the quasar in the past has been responsible for cooling the ICM in the central 80 kpc of the cluster, thus lowering the gas temperature and entropy. The timescale for Compton cooling, $t_{C}$ depends only on the luminosity of the quasar and the distance from it (\citealt{Fabian1990});
\begin{equation}
t_{C} = 10^{13} R_{2}^{2} L_{47}^{-1} {\rm s}
\label{tcompton}
\end{equation} 
where $R_{2}$ is the radial distance in units of 100 parsecs, and $L_{47}$ is the luminosity of the quasar in units of 10$^{47}$ ergs s$^{-1}$. 
Extrapolating the temperature and density profiles inwards from the 8-20 arcsecond region, we find that the current quasar bolometric luminosity of 2$\times$10$^{47}$ ergs s$^{-1}$ means that Compton cooling dominates over bremsstrahlung cooling within the central 2 kpc, within which the gas would be expected to rapidly cool down to the Compton temperature. We compare with bremsstrahlung cooling, as this will dominate in the highly ionised conditions near to the quasar (\citealt{Fabian1990}). Using the quasar spectral energy distribution from the optical to the soft X-ray, \citet{Russell2010} calculated that the Compton temperature is $T_{C}\sim0.4$ keV. In order for Compton cooling to have been the dominant cooling mechanism out to 80 kpc at some stage in the past, we have calculated that the bolometric luminosity of the quasar would have needed to be 20 times higher, 4$\times$10$^{48}$ ergs s$^{-1}$, corresponding to the Eddington limit of a 3$\times$10$^{10}$ M$_{\odot}$ black hole. This is at the extreme end of what is realistically possible, and this scenario would seem unlikely.

This black hole mass is similar to the most massive black holes recently inferred in \citet{Larrondo2012} which are necessary for the black holes in the brightest cluster galaxies (BCGs) of massive cool core clusters to lie on the fundamental plane of black hole activity. Highly massive, $10^{10}$ M$_{\odot}$, black holes have been measured in   the BCG's of the Coma cluster and Abell 1367 (NGC 4889 and NGC 3842 respectively) in \citet{McConnell2011}, which are significantly more massive that would be predicted by linearly extrapolating commonly used scaling relations between black hole mass and the host galaxy bulge luminosity or stellar velocity dispersion. It is not possible at present to accurately measure the black hole mass of H1821+643. In \citet{Russell2010} it was estimated that the black hole mass was $\sim$3$\times$10$^{9}$ M$_{\odot}$, though due to the number of extrapolations involved this was just a rough estimate. The required underestimate of the mass is at least log($\Delta M_{BH}$) = 1.0, which is reasonably consistent with the average underestimate of black hole mass found in \citet{Larrondo2012} of  log($\Delta M_{BH}$)=0.8$\pm$0.6 which would be necessary to bring the black holes in BCGs into agreement with the fundamental plane. We therefore conclude that Compton cooling from the quasar radiation in a previously brighter state may provide a possible solution for the low temperature and entropy observed in the central 80kpc of the cluster surrounding H1821+643, but the extreme luminosity and black hole mass necessary make this scenario unlikely.

A higher black hole mass would, however, make it easier to explain the current observed luminosity of the quasar. \citet{Russell2010} showed that the accretion rate needed to power the quasar in its current state, with a bolometric luminosity of 2$\times$10$^{47}$ ergs s$^{-1}$ is 40 M$_{\odot}$ yr$^{-1}$ for a radiative efficiency of 0.1. However the Bondi accretion rate (using the innermost temperature and density measurements) was calculated to be only 0.001 M$_{\odot}$ yr$^{-1}$. This can rise to $\sim$6 M$_{\odot}$ yr$^{-1}$ if the accreting material has been Compton cooled within 2 kpc, (with the Bondi radius being $\sim$250 pc), where Compton cooling should dominate bremsstrahlung cooling for the current quasar luminosity, but is still nearly an order of magnitude below the accretion rate needed to power the quasar at present. If the black hole was more massive than the value obtained in \citet{Russell2010}, this would increase the Bondi accretion rate and make it easier to explain the current observed quasar luminosity. However the order of magnitude increase from $\sim$3$\times$10$^{9}$ M$_{\odot}$ to $\sim$3$\times$10$^{10}$ M$_{\odot}$ would seem to increase the accretion rate too much (it would be $\sim$600 M$_{\odot}$ yr$^{-1}$). 

\subsection{Comparison to a constant pressure cooling flow}
\label{cooling_flow_comparison}
\begin{figure}
  \begin{center}
    \leavevmode
    \vbox{
      \epsfig{figure=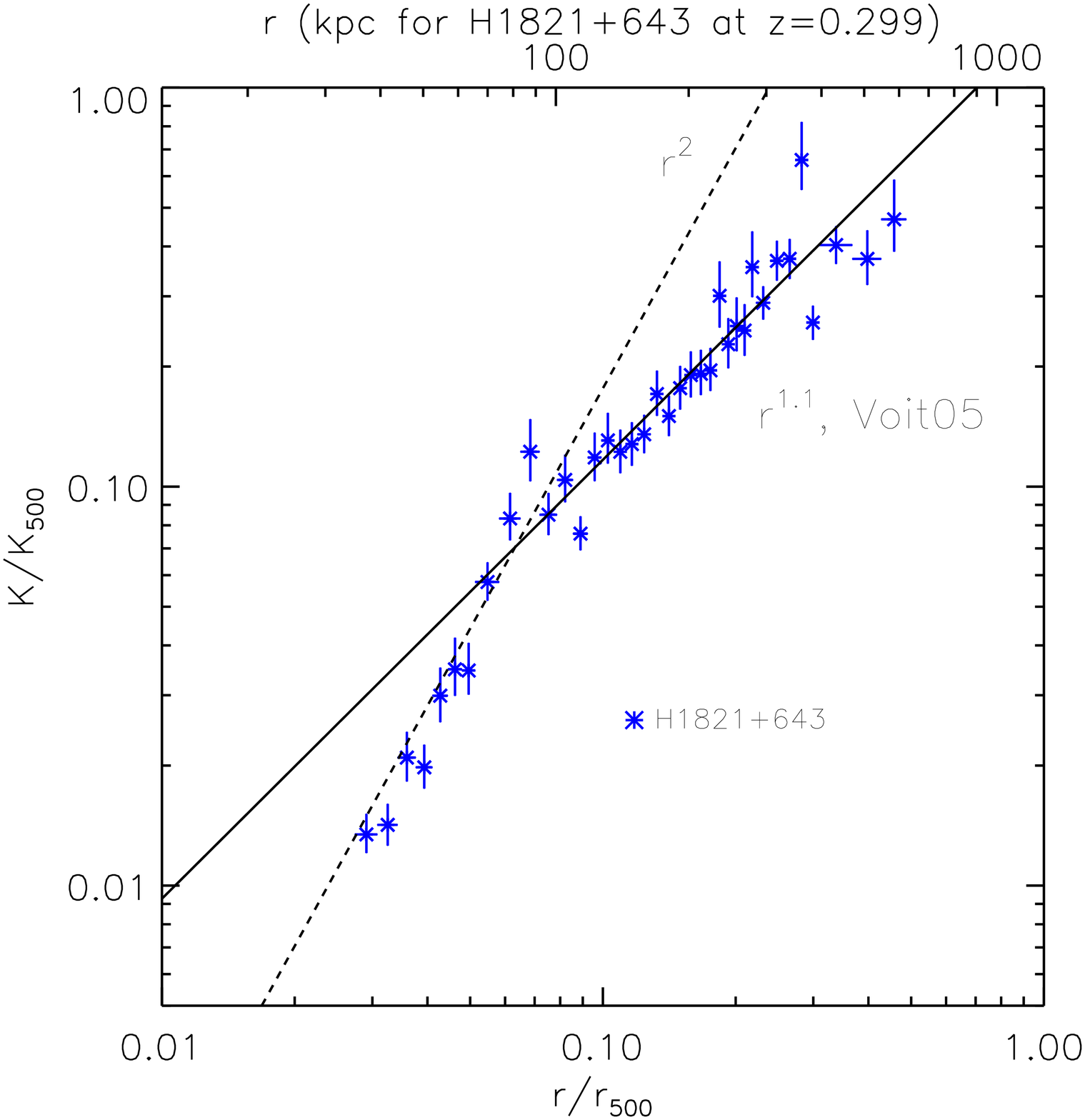,
        width=0.9\linewidth}
         \epsfig{figure=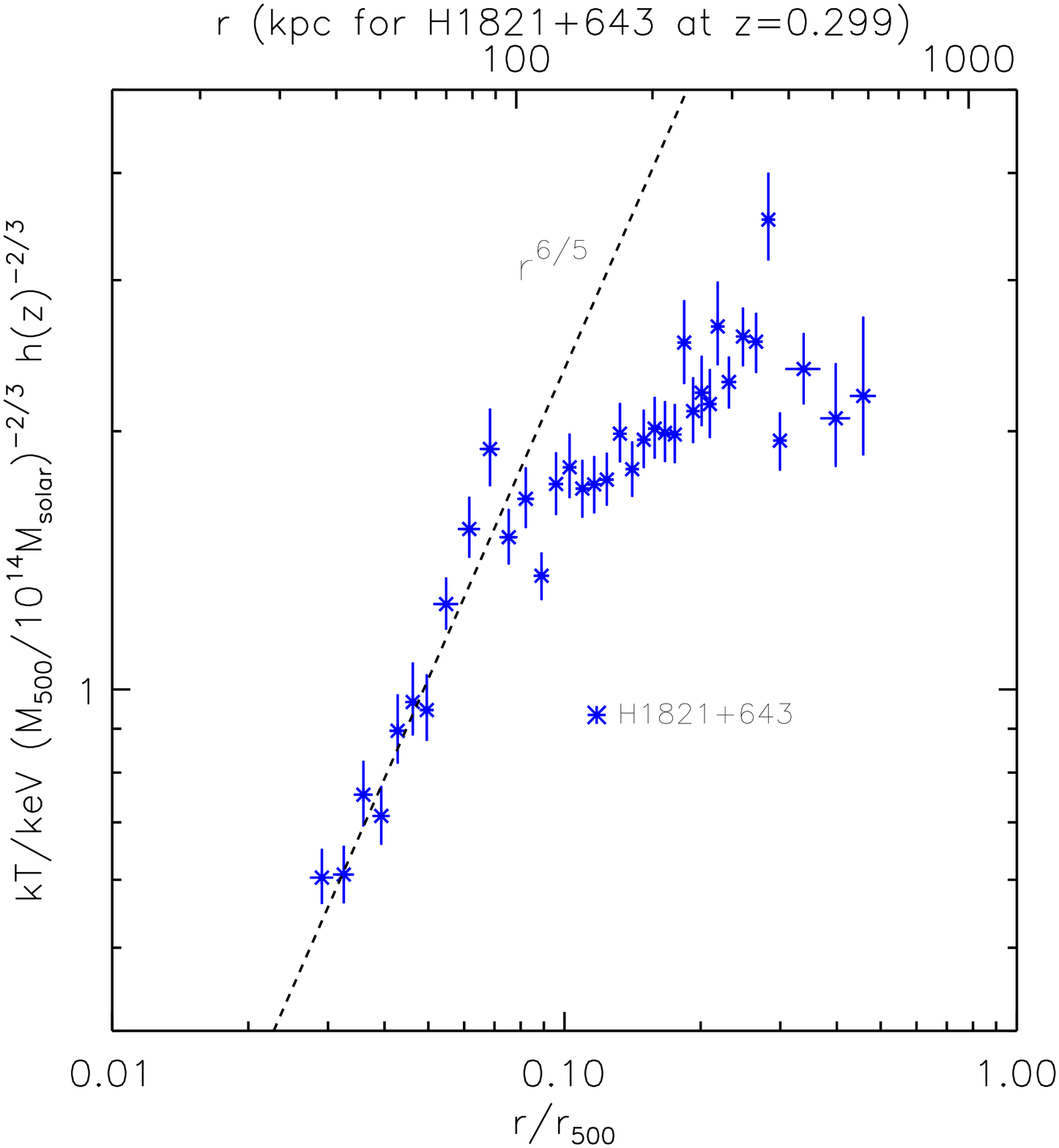,
        width=0.9\linewidth}    
                  \epsfig{figure=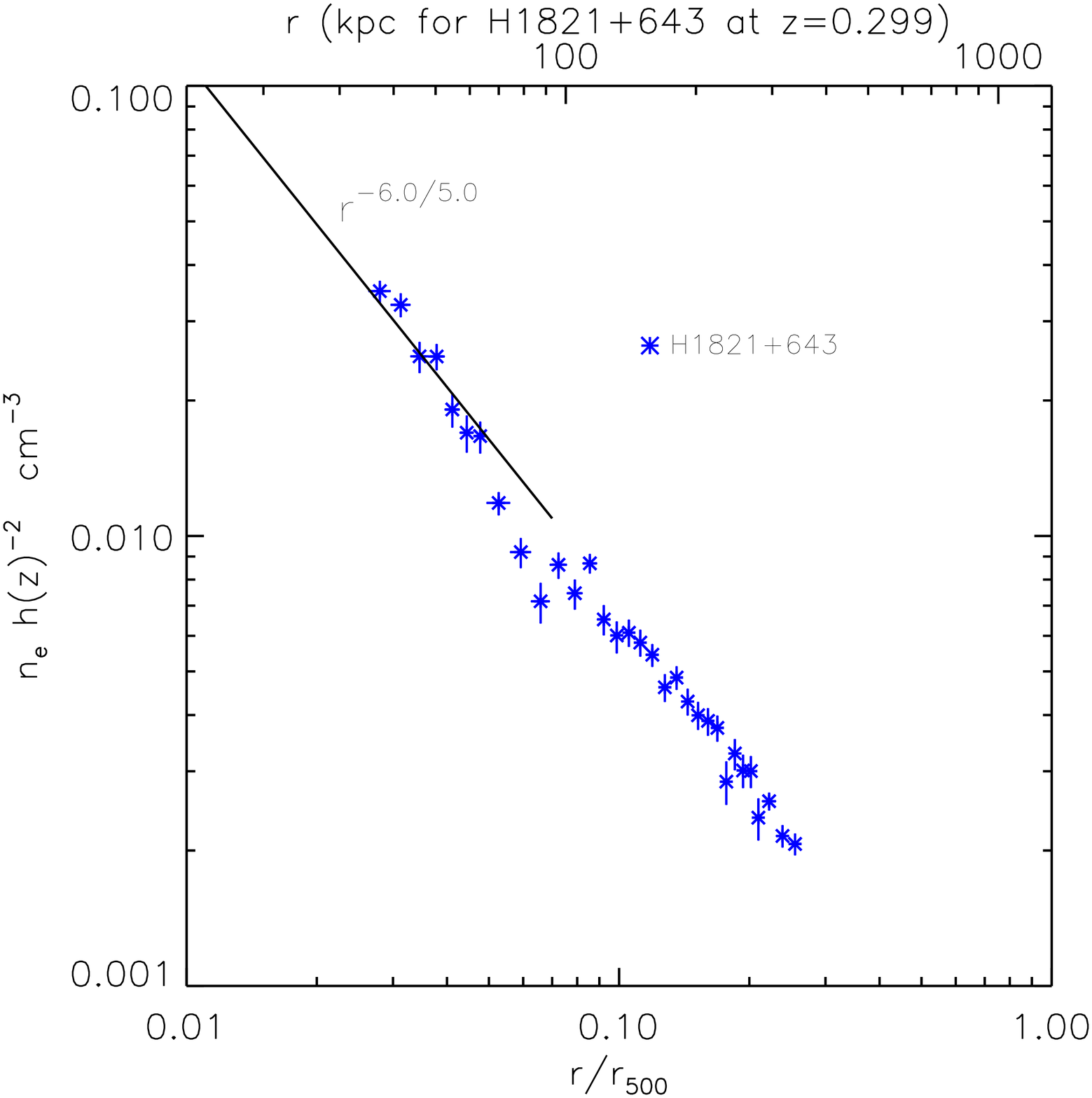,
        width=0.9\linewidth}
        }
   
      \caption{Comparing the central entropy, temperature and density profiles shapes to the predictions of a constant pressure cooling flow model.    }
      \label{cf_profiles}
  \end{center}
\end{figure}

For a cooling flow where the mass flux is fixed, the flow is subsonic and the pressure is nearly constant (so that the flow is driven by pressure and cooling and not gravity), \citet{Nulsen1982} derived that the density and temperature profiles should vary as (see also \citealt{Fabian1984}, \citealt{Sarazin1988});
\begin{equation}
\rho_{g} \propto 1/T_{g} \propto r^{-3/(3- \alpha )}
\label{cooling_flow1}
\end{equation}
where $\alpha$ is the powerlaw dependence of the cooling function on the temperature, $\Lambda \propto T_{g}^{\alpha}$. When thermal bremsstrahlung cooling dominates, $\alpha$=0.5, yielding;
\begin{equation}
\rho_{g} \propto r^{-6/5} 
\label{cooling_flow2}
\end{equation}
\begin{equation}
T_{g} \propto r^{6/5 }
\label{cooling_flow3}
\end{equation}
so that the entropy profile will vary as
\begin{equation}
K \propto T/n^{2/3} \propto r^{6/5 }/ r^{-6/5 \times 2/3} \propto r^{2}
\label{cooling_flow4}
\end{equation}
In Fig. \ref{cf_profiles} we compare the entropy, temperature and density profile shapes with these cooling flow profile shapes. Fascinatingly, the temperature and entropy profiles seem to be in reasonable agreement with the predicted profiles. 

We fit the deprojected spectrum (using \textsc{dsdeproj} to perform the deprojection) in the 30-80kpc region with the cooling flow model MKCFLOW (\citealt{Mushotzky1988}). We find a good fit (a reduced $\chi^2$ of 1.07 for 177 degrees of freedom) with sensible best fitting results when we fix the upper temperature to 8 keV, and allow the lower temperature and mass deposition rate to be free parameters. We find a lower temperature of 2.5$^{+0.6}_{-0.5}$ keV and mass deposition rate of 630$^{+99}_{-62}$ M$_{\odot}$ yr$^{-1}$. 

In order for such a cooling flow to form, it is necessary that some mechanism prevents feedback from the quasar from heating the gas in the cooling flow and raising its entropy. The radio data indicate that the jets from the quasar extend at least 70 kpc from the core to the north and south (see figure 4 of \citealt{Russell2010}), and likely reach out to a radius of around 100 kpc when projection effects are taken into account. It is therefore possible that the feedback energy from these jets is dissipated around 100 kpc from the centre of the cluster and has little effect inside this radius, thus allowing the cooling flow to exist within the central 80 kpc.  

To see whether the jets have a significant impact on the temperature, density and entropy profiles, we restricted the aperture to study regions only along the jets in 60 degree angle sectors to the north and south. The temperature, entropy and density profiles were obtained using the surface brightness deprojection method. We then found profiles by restricting the aperture to the regions away from the jets to the east and west along sectors with opening angles of 60 degrees. These sectors are shown in white on Fig. \ref{image}. These profiles are plotted in Fig. \ref{compare_directions} in Appendix \ref{sec:appendix}. We see that there are no significant differences between the profiles along the jets (black points) and away from the jets (red points). 

The highly energetic environment around the central quasar, which includes powerful jets, makes this system fundamentally different to other strong cool core clusters, and may prevent the accumulation of reservoirs of molecular gas. Measurements of the CO emission in H1821+643 (\citealt{Aravena2011}) indicate that there is $\sim$8.0$\times$10$^{9}$ M$_{\odot}$ of molecular gas located $\sim$9 kpc southeast from the nucleus position. This CO emission is roughly coincident with and elongated along the quasar jet axis, and may be interacting.   

%In Fig. \ref{cooling_rate} we show profiles of the integrated mass cooling rate and the cooling time. 

%\begin{figure}
%  \begin{center}
%    \leavevmode
%    \vbox{
%      \epsfig{figure=cooling_rateandtime.eps,
%        width=0.9\linewidth}
% 
%        }
%   
%      \caption{Plots of the profiles of the integrated mass cooling rate and the cooling time.     }
%      \label{cooling_rate}
%  \end{center}
%\end{figure}

\begin{figure}
  \begin{center}
    \leavevmode
    \vbox{
      \epsfig{figure=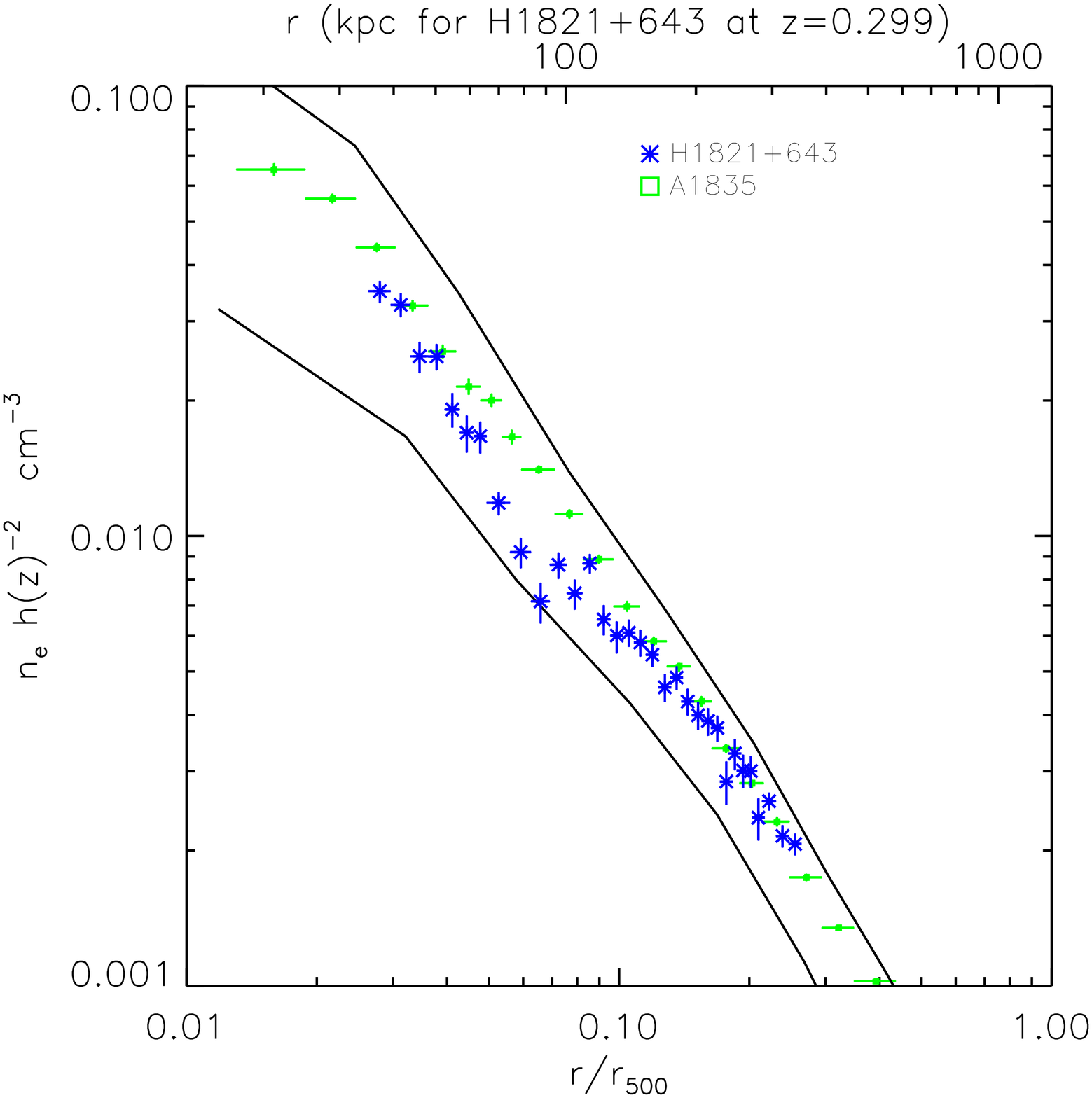,
        width=0.9\linewidth}
 
        }
   
      \caption{Comparing the density profile of H1821+643 with that of the similarly massive A1835, with the expected redshift evolution scaled out. The solid black lines show the bounds of the range of the density profiles for strong cool core clusters found in \citet{Morandi2007}.      }
      \label{dprofilecompare}
  \end{center}
\end{figure}

If gas has been cooled out and accreted by the quasar, we would expect this to have had an effect on the gas density profile. In Fig. \ref{dprofilecompare}, we compare the gas density profile to that of Abell 1835, which has essentially the same mass and lies at roughly the same redshift as H1821+643. We see that the gas density profile is significantly lower than that of Abell 1835 in the 0.03-0.09$r_{500}$ range, and that the missing gas starts at the radius where the temperature and entropy profiles break. Away from the core the density profiles tend towards the same self similar profile as expected (e.g. \citealt{Croston2008}). However, since the intrinsic scatter for the density profiles of strong cool core clusters is around a factor of 2 in the 0.03-0.09$r_{500}$ range (the solid black lines on Fig. \ref{dprofilecompare} show the range found for the 6 strong cool core clusters examined in \citealt{Morandi2007} for example), the density profiles are in reasonable agreement given this expected scatter. It therefore is not possible to make an accurate calculation for the difference in gas mass, or to say for certain that the ICM around H1821+643 is gas deficient.  If the difference between the density profiles of A1835 and H1821+643 is converted into gas mass, which can be interpreted as an upper limit on the mass which could have accreted onto the quasar, the result is colossal (10$^{12}$ M$_{\odot}$). This shows that there is comfortably enough gas that could have accreted onto the quasar to form a black hole of mass $\sim$3$\times$10$^{10}$ M$_{\odot}$.

\section{Conclusions}  
We have found that the intracluster medium around the quasar H1821+643 is significantly different to that found in other galaxy clusters. The entropy and temperature are significantly lower, and have much steeper gradients, than is observed in other massive cool core clusters. The entropy profile within 80kpc from the core lies significantly below the baseline entropy profile for hierarchical structure formation, a phenomenon which has never been observed before, and indicates that the ICM around the quasar has been significantly cooled. Fascinatingly, the gradients of the temperature and entropy profiles within 80 kpc are in good agreement with the expected profiles for a constant pressure cooling flow. 

One possible way in which the quasar could have cooled the ICM is through Compton cooling, though for this influence to extend to 80 kpc from the core, we would require that the quasar has been at least 20 times as luminous as it is today, corresponding to the Eddington luminosity of an incredibly massive $\sim$3$\times$10$^{10}$ M$_{\odot}$ black hole, similar to the most massive black holes recently suggested in the analysis of \citet{Larrondo2012}. 

It is possible that the system has been locked into a Compton cooled feedback cycle which prevents energy release from the black hole heating the gas sufficiently to switch it off, leading to the formation of a huge ($\sim$3$\times$10$^{10}$ M$_{\odot}$) supermassive black hole. If Compton cooling dominates near the centre of clusters around luminous quasars, then
it can stop heat propagating out through the gas, thus allowing a regular cooling flow to exist. In the case of H1821+643, which has jets extending out to around 70-100 kpc from the cluster centre when projection effects are taken into account, it is possible that the feedback energy from the jets is dissipated in the ICM in the 80-100 kpc range, and has little effect on the ICM inside this radius, thus allowing the gas the cool in the central 80kpc to the low temperature and entropy observed. Such a mechanism may be common around the quasars located in clusters at higher redshifts, and may provide a way of forming massive, $\sim$ $10^{10}$ M$_{\odot}$, black holes in the cluster BCG's, which the analysis of \citet{Larrondo2012} suggests is necessary to allow these black holes to lie on the fundamental plane of black hole activity. 

We note that in the Phoenix cluster, (\citealt{McDonald2012}, \citealt{McDonald2013}), which has the highest known mass deposition rate of all strong cool core clusters (2700$\pm$700 M$_{\odot}$ yr$^{-1}$ when the gravitational work done is taken into account, with a star formation rate which is 30 percent of this), there is a very luminous AGN, which although absorbed along our line of sight, may not be absorbed in all directions. The mass of the supermassive black hole in the central galaxy of the Phoenix cluster is estimated in \citet{McDonald2012} to be extremely massive, $\sim$ 1.8$^{+2.5}_{-1.2}\times$10$^{10}$ M$_{\odot}$. It is possible that a Compton cooled feedback cycle has played a role in forming this cooling flow and growing this supermassive black hole during a past quasar phase. If Compton cooling still dominates near the core then this could prevent the AGN from heating the gas and counteract the cooling, thus allowing a strong cooling flow to continue to exist.

%3620$\pm$530 M$_{\odot}$ yr$^{-1}$, with a star formation rate which is 20 percent of this
%\section{Summary}
%
%
%    
%
%\label{summary}

\section*{Acknowledgements}

SAW is supported by the ERC. This
work is based on observations obtained with the \emph{Chandra} observatory, a NASA mission.

\bibliographystyle{mn2e}
\bibliography{Quasar_paper}

\appendix

\section[]{}

\label{sec:appendix}

\begin{figure}
  \begin{center}
    \leavevmode
    \vbox{
      \epsfig{figure=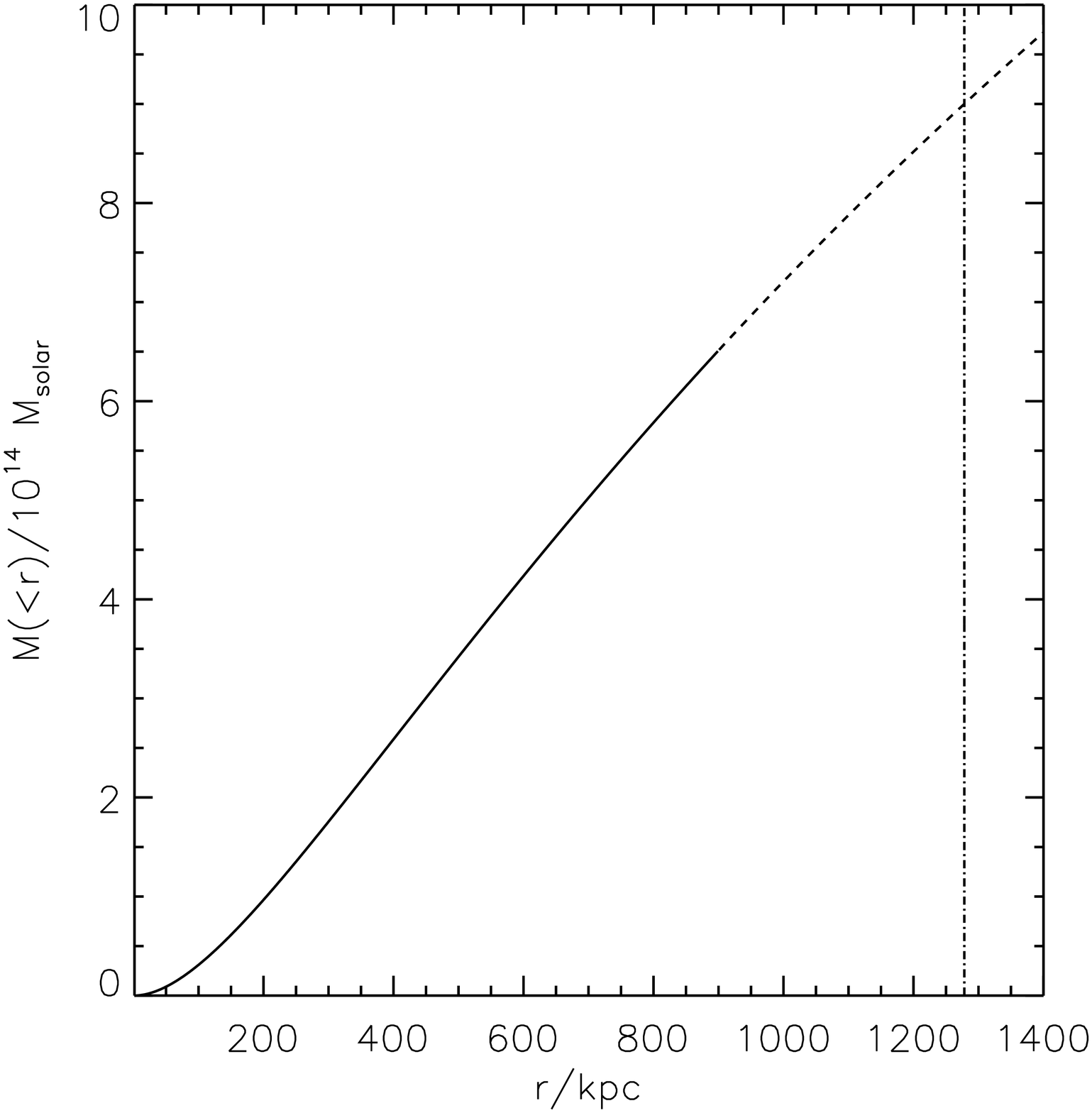,
        width=0.9\linewidth}
 
        }
   
      \caption{Best fit total NFW mass profile for H1821+643 determined through the method of \citet{Schmidt2007}, which uses temperature measurements within the region where the mass profile is solid (the central 900kpc). Outside 900 kpc it is not possible to measure the ICM temperature, and the best fit NFW profile is extrapolated outwards to find $M_{500}$ (the dashed section of the line). The vertical dot-dash line shows the position of $r_{500}$.       }
      \label{massprofile}
  \end{center}
\end{figure}

\begin{figure}
  \begin{center}
    \leavevmode
    \vbox{
      \epsfig{figure=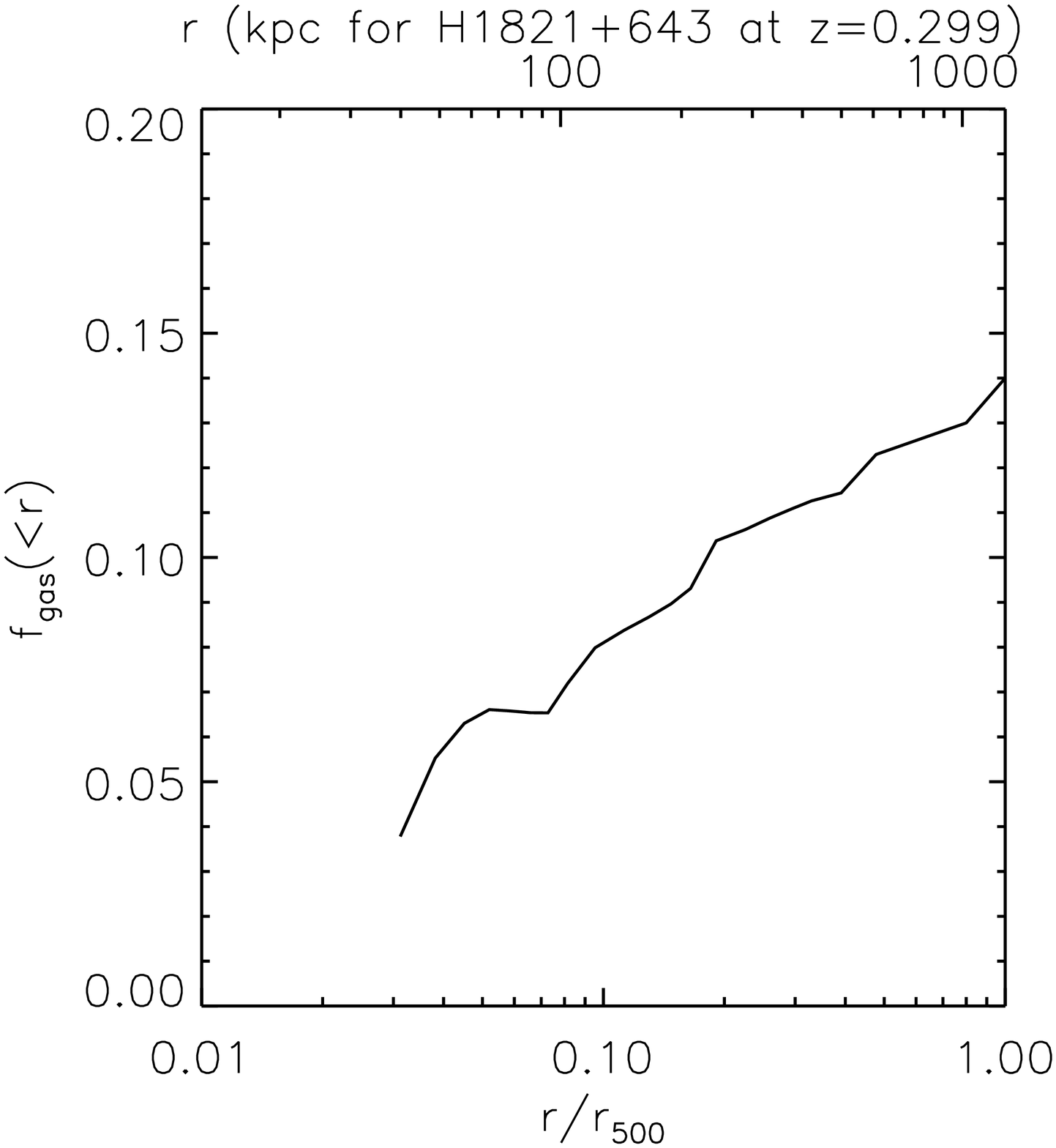,
        width=0.9\linewidth}
 
        }
   
      \caption{Gas mass fraction profile for H1821+643. It is not possible to measure $f_{gas}$ in the central 30 kpc due to contamination from the quasar.       }
      \label{fgasprofile}
  \end{center}
\end{figure}

\begin{figure}
  \begin{center}
    \leavevmode
    \vbox{
      \epsfig{figure=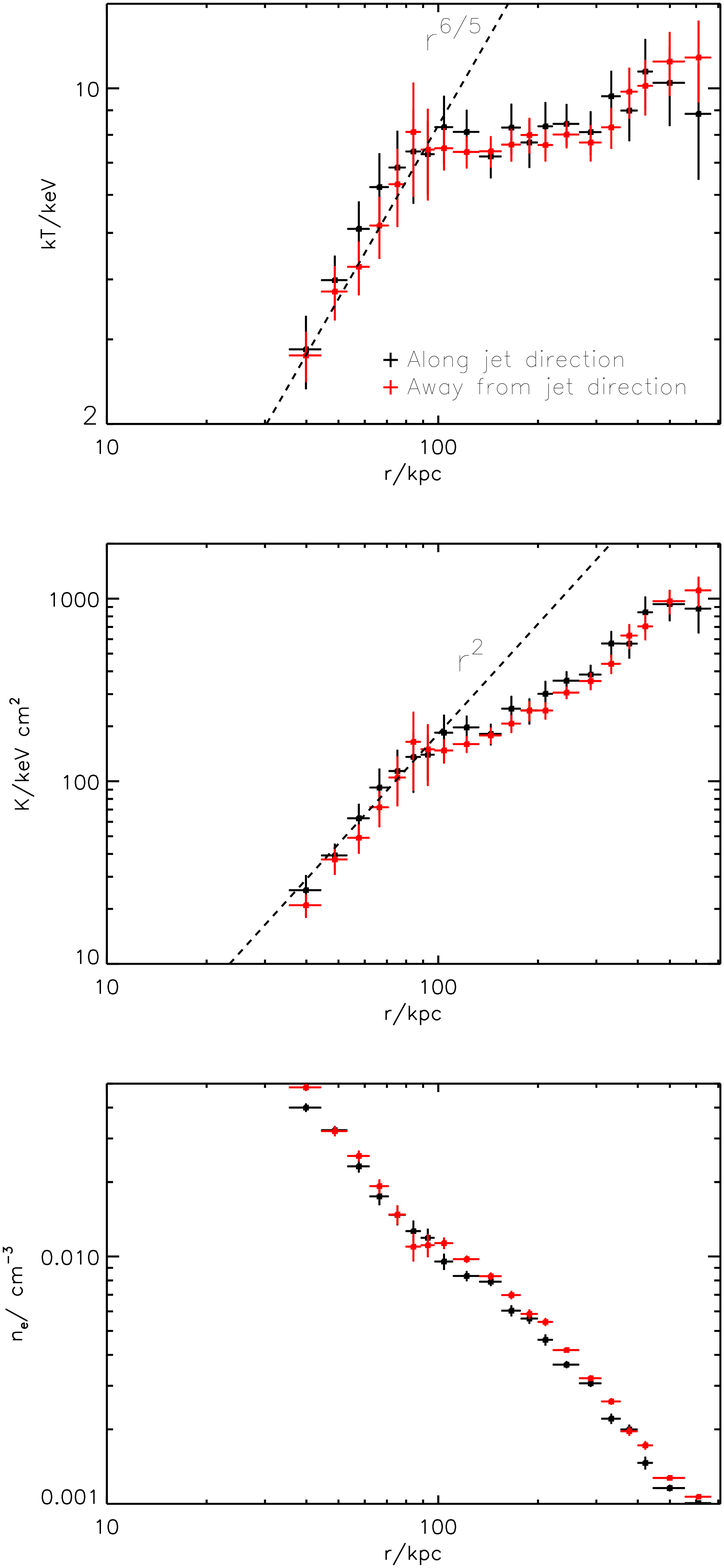,
        width=0.9\linewidth}
 
        }
   
      \caption{Comparing the temperature (top), entropy (middle) and density (bottom) profiles for directions along the jets (black points) and away from the jets (red points), both of which agree with one another well.       }
      \label{compare_directions}
  \end{center}
\end{figure}

Fig. \ref{massprofile} shows the total mass profile for H1821+643, while Fig. \ref{fgasprofile} shows the gas mass fraction profile. Fig. \ref{compare_directions} compares the temperature, entropy and density profiles along the jet directions with those away from the jet directions. \newline

%
%
%
%
%
%
%\clearpage

%\section[]{ROSAT imaging in E11}

\end{document}